\documentclass[12pt,a4paper,twoside,english]{article}
\usepackage[T1]{fontenc}
\usepackage[latin1]{inputenc}
\usepackage{color}
\usepackage{amssymb}
\usepackage{graphicx}

\makeatletter

\newcommand{\noun}[1]{\textsc{#1}}

\makeatother

\usepackage{babel}
\begin{document}
\title{\textcolor{black}{Mechanism of production and deviation from the standard
Gutenberg-Richter law of the big earthquakes }\\
\textcolor{black}{(An analysis of big earthquakes)}}
\author{\textcolor{black}{\normalsize{}B. F. Apostol }\\
\textcolor{black}{\normalsize{}National Institute for Earth's Physics,
Magurele-Bucharest MG-6, }\\
\textcolor{black}{\normalsize{}POBox MG-35, Romania }\\
\textcolor{black}{\normalsize{}email: afelix@theory.nipne.ro}}
\date{\textcolor{black}{{}}}

\maketitle
\textcolor{black}{Orchid: 0000-0002-9990-9390}

\textcolor{black}{\relax}
\begin{abstract}
\textcolor{black}{We describe two self-replicating mechanisms of energy
accumulation in the seismic focus, which modify the Gutenberg-Richter
law in the region of the big earthquakes. The first mechanism acts
for magnitudes smaller than a narrow region of large critical magnitudes;
it slows down the energy accumulation and may produce precursors.
The second mechanism acts above that region, and accelerates the energy
accumulation; the precursors may be absent. Both mechanisms reduce
the Gutenberg-Richter excedence distribution. On the left of the critical
region the Gutenberg-Richter magnitude probability density is unchanged,
while on the right the probability density is reduced. The procedure
described in this paper introduces a critical-magnitude region (range)
as an additional fitting parameter. The results may bear relevance
upon the recent concepts of \textquotedbl self-arresting\textquotedbl{}
and \textquotedbl dragon-king\textquotedbl{} earthquakes. The two self-replicating
mechanisms may introduce a magnitude gap between the two types of
big earthquakes, and two branches in the excedence law, in the vicinity
of the critical region.}
\end{abstract}
\textcolor{black}{\relax}

\textcolor{black}{Key words: Gutenberg-Richter distributions; excedence
law; big earthquakes; self-replication}

\textcolor{black}{It is well known that one of the basic empirical
laws in seismology is the Gutenberg-Richter law, which states that
the fraction of the earthquakes with (moment) magnitude greater than
$M$ (excedence, cummulative frequency) is given by 
\begin{equation}
P=e^{-\beta M}\,\,.\label{1}
\end{equation}
The fitting parameter $\beta$ is known as the Gutenberg-Richter parameter;
$\beta\simeq2.3$ ($1$ for powers of ten) is usually accepted as
a reference value (Stein \& Wysession, 2003; Udias, 1999; Lay \& Wallace,
1995; Frohlich \& Davis, 1993). The law applies to a given seismic
region where we have $N$ earthquakes with magnitude greater than
$M$, out of a total number $N_{0}$, occurring in a long time $T$,
such that we may write $P=N/N_{0}$ and $N_{0}=T/t_{0}$, where the
seismicity rate $1/t_{0}$ is another fitting parameter; consequently,
from equation (\ref{1}) we get the well-known logarithmic law
\begin{equation}
\ln(N/T)=-\ln t_{0}-\beta M\,\,.\label{2}
\end{equation}
}

\textcolor{black}{Also, from equation (\ref{1}) we get the Gutenberg-Richter
probability 
\begin{equation}
dP_{m}=-\frac{dP}{dM}dM=\beta e^{-\beta M}dM\label{3}
\end{equation}
of having an earthquake with magnitude $M$ in the interval $(M,M+dM)$,
and its logarithmic form 
\begin{equation}
\ln(dP_{m}/dM)=\ln\beta-\beta M\,\,.\label{4}
\end{equation}
These laws are well documented for statistical ensembles of earthquakes,
except for low magnitudes, where the straight lines given by equations
(\ref{2}) and (\ref{4}) become flattened (the so-called roll-off
effect, Bhattacharya et al., 2009; Pelletier, 2000; Jones, 1994),
and for big earthquakes, where interesting deviations may appear (Main,
1992; Wesnousky, 1994; Hamilton \& McCloskey, 1997; Sammis \& Sornette,
2002; Pisarenko \& Sornette, 2004; Saichev \& Sornette, 2006, 2007;
Ishibe \& Shimazaki, 2012). Usually, the deviations occurring in the
region of large magnitudes are assigned to insufficient data, catalog
incompleteness, saturation effects, etc. On the other hand, it may
sound reasonable to investigate whether such deviations are due to
physical causes. Often, the earthquakes distribution in this region
looks as if there exists a large critical magnitude $M_{c}$, around
which the earthquake frequency is smaller than that predicted by the
Gutenberg-Richter law, with two distinct branches for }\textcolor{black}{\emph{$M<M_{c}$
}}\textcolor{black}{and $M>M_{c}$. Sometimes the first group ($M<M_{c}$)
is called \textquotedbl arrested\textquotedbl{} (\textquotedbl self-arresting\textquotedbl )
earthquakes (Sornette et al., 2026) and the second group ($M>M_{c}$)
\textquotedbl dragon-king\textquotedbl{} earthquakes (Sornette, 2009),
though such denominations, as well as the existence of a parameter
$M_{c}$, separating two distinct regimes, do not enjoy consensus.
These concepts were analyzed recently in the context of nucleation
conditions, fault dynamics, stress evolution, etc (Li et al, 2024;
Sornette et al., 2026).}

\textcolor{black}{We describe in this paper two distinct mechanisms
which may explain the existence of a critical region around a parameter
$M_{c}$, and two distinct, adjoining distribution branches. The large
anomalies of the big earthquakes in this region may be interpreted
in terms of these two distribution branches, arising from two distinct
self-replication mechanisms. }
\begin{figure}
\begin{centering}
\textcolor{black}{\includegraphics[clip,scale=0.8]{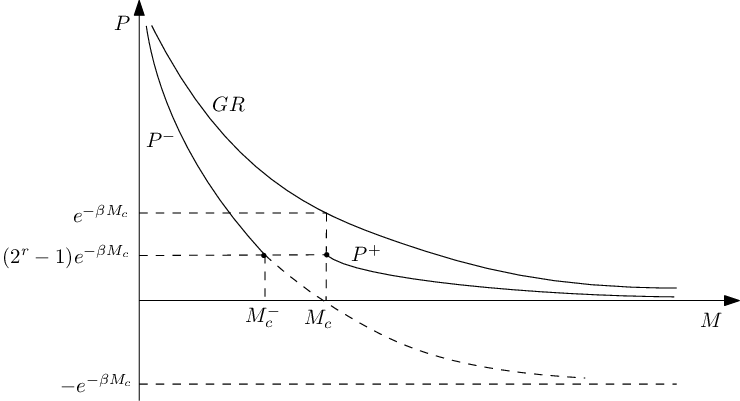}}
\par\end{centering}
\textcolor{black}{\caption{The excedence frequency $P$ given by equation (\ref{25}) vs magnitude
$M$, with the two branches $P^{(-,+)}$ separated by the critical
magnitude $M_{c}$, compared with the Gutenberg-Richter excedence
frequency (GR). \label{fig:Fig.1}}
}
\end{figure}

\textcolor{black}{In Apostol (2006a) the law 
\begin{equation}
t=t_{0}(E/E_{0})^{r}\label{5}
\end{equation}
 of accumulating an energy $E$ in time $t$ in a pointlike seismic
focus has been established, where $E_{0}$ is a small cutoff energy
and $r$ is a parameter which lies in the range $1/3<r<1$; very likely,
the reference value is around $r=2/3$. By using the definition of
the magnitude }

\textcolor{black}{
\begin{equation}
E=E_{0}e^{bM}\,\,\,,\label{6}
\end{equation}
 (Utsu \& Seiki 1955; Utsu, 1969; Kanamori, 1977; Hanks \& Kanamori,
1979; Gutenberg \& Richter, 1944, 1956), where $b=3.45$ ($3/2$ for
powers of ten), we get from equation (\ref{5}) the mean recurrence
time 
\begin{equation}
t=t_{0}e^{\beta M}\,\,\,,\label{7}
\end{equation}
where $\beta=rb$. The frequency of the fundamental seismic events
with energy $E_{0}$ in time $t$, each during a short time $t_{0}$,
viewed as independent events, is $t_{0}/t$. This is also the generating
function of the Gutenberg-Richter distribution 
\begin{equation}
dP_{m}=-\frac{d}{dt}\frac{t_{0}}{t}dt=\beta e^{-\beta M}dM\label{8}
\end{equation}
for independent earthquakes, which is equation (\ref{3}); also, the
excedence law for such events is 
\begin{equation}
P=\int_{t}^{\infty}dP_{m}=\frac{t_{0}}{t}=e^{-\beta M}\,\,\,,\label{9}
\end{equation}
as in equation (\ref{1}).}

\textcolor{black}{We describe below two self-replication processes
of energy accumulation. The mechanism of self-replication is a repetitive
process indicated by equation (\ref{5}), with different rates. When
the energy accumulated is too large, the process slows down, while
for a long accumulation time the process accelerates. These two distinct
rates lead to finite amounts of (large) energies in finite acumulation
times. }

\textcolor{black}{Let us introduce a large critical energy $E_{c}$,
corresponding to the large critical magnitude $M_{c}$ given by $E_{c}=E_{0}e^{bM_{c}}$
and the long critical time $t_{c}=t_{0}(E_{c}/E_{0})^{r}$.}

\textcolor{black}{We write the accumulation time as }

\textcolor{black}{
\begin{equation}
\begin{array}{c}
t=t_{c}+t_{0}(E_{c}/E_{0})^{r}(E/E_{c})^{r}-t_{c}=\\
\\
=t_{0}(E_{c}/E_{0})^{r}\left[1+(E/E_{c})^{r}-1\right]\,\,.
\end{array}\label{10}
\end{equation}
 In this equation we may view the first two terms in the bracket as
indicative of a self-replication process (Apostol, 2006b),
\begin{equation}
\begin{array}{c}
1+(E/E_{c})^{r}+...=1+(E/E_{c})^{r}\left[1+(E/E_{c})^{r}+...\right]=\\
\\
=\frac{1}{1-(E/E_{c})^{r}}\,\,\,,
\end{array}\label{11}
\end{equation}
such that equation (\ref{10}) becomes
\begin{equation}
t=t_{0}(E_{c}/E_{0})^{r}\left[\frac{1}{1-(E/E_{c})^{r}}-1\right]\,\,.\label{12}
\end{equation}
}

\textcolor{black}{In order to accumulate an energy $E$ we need a
time $t_{0}(E/E_{0})^{r}$, which we write as $t_{0}(E_{c}/E_{0})^{r}(E/E_{c})^{r}$,
}\textcolor{black}{\emph{i.e.}}\textcolor{black}{{} we reduce the time
of accumulating the threshold energy $E_{c}>E$ by fraction $(E/E_{c})^{r}$
(reduction rate). This reduction process may act again, and we add
$t_{0}(E_{c}/E_{0})^{r}(E/E_{c})^{2r}$, $t_{0}(E_{c}/E_{0})^{r}(E/E_{c})^{3r}$,
...., in a repetitive process. This is the self-replication process
described above. The process of energy accumulation tends to achieve
the energy $E_{c}$, but, for various reasons (for example energy
loss, dissipation, etc), it is slowed down to some energy $E<E_{c}$. }

\textcolor{black}{We can see that $E<E_{c}$, and there exists a singularity
at $E=E_{c}$. This self-replication process requires an infinite
time to reach the critical energy $E_{c}$; it is a slowed down process
of energy accumulation. Of course, the region of $E$ very close to
$E_{c}$ is unphysical. The excedence frequency given by equation
(\ref{12}) is
\begin{equation}
P^{-}=\frac{t_{0}}{t}=(E_{0}/E)^{r}-(E_{0}/E_{c})^{r}\,\,\,,\label{13}
\end{equation}
or 
\begin{equation}
P^{-}=e^{-\beta M}-e^{-\beta M_{c}}\,\,;\label{14}
\end{equation}
 the superscript \textquotedbl$-$\textquotedbl{} indicates the region
$M<M_{c}$. From equation (\ref{14}) we can see that for small and
moderate magnitudes the Gutenberg-Richter excedence law is practically
unchanged ($e^{-\beta M_{c}}$ may be omitted in this region), while
it decreases to zero for $M\rightarrow M_{c}$. The Gutenberg-Richter
probability density is unchanged, 
\begin{equation}
-dP^{-}/dM=\beta e^{-\beta M}\,\,\,,\label{15}
\end{equation}
except for the fact that it is limited to $M<M_{c}$; this result
may correspond to \textquotedbl self-arresting\textquotedbl{} earthquakes.
}
\begin{figure}
\begin{centering}
\textcolor{black}{\includegraphics[clip,scale=0.8]{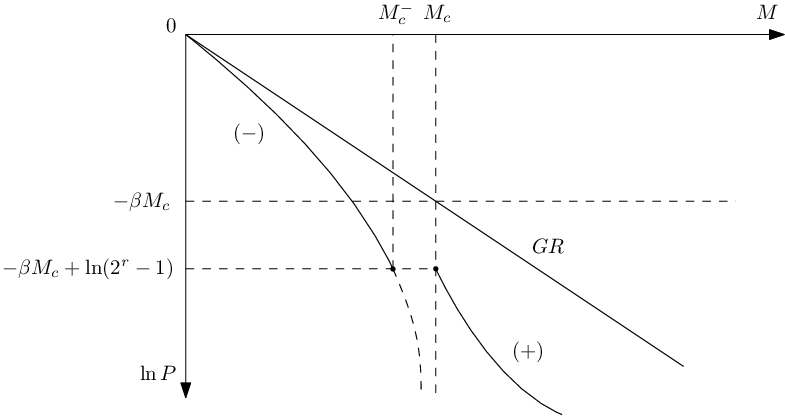}}
\par\end{centering}
\textcolor{black}{\caption{The logarithmic form of the excedence frequ\textcolor{red}{ency giv}en
in Fig. \ref{fig:Fig.1}. \label{fig:Fig.2}}
}
\end{figure}
\textcolor{black}{{} Since the energy accumulation for the big earthquake
is delayed in this case, we may admit that the energy excess (if any)
is taken up by precursor, smaller earthquakes.}

\textcolor{black}{We show now that energies larger than $E_{c}$ (}\textcolor{black}{\noun{$M>M_{c}$)
}}\textcolor{black}{can be attained by an accelerated process of energy
accumulation. To this end we write 
\begin{equation}
\begin{array}{c}
E=E_{c}+E_{0}(t_{c}/t_{0})^{1/r}(t/t_{c})^{1/r}-E_{c}=\\
\\
=E_{0}(t_{c}/t_{0})^{1/r}\left[1+(t/t_{c})^{1/r}-1\right]
\end{array}\label{16}
\end{equation}
and, by the same self-replication procedure as above, 
\begin{equation}
E=E_{0}(t_{c}/t_{0})^{1/r}\left[\frac{1}{1-(t/t_{c})^{1/r}}-1\right]\,\,.\label{17}
\end{equation}
The energy accumulated in time $t$ is $E=E_{0}(t/t_{0})^{1/r}$,
which we write as $E_{0}(t_{c}/t_{0})^{1/r}(t/t_{c})^{1/r}$, }\textcolor{black}{\emph{i.e.}}\textcolor{black}{{}
the threshold time $t_{c}$ is reduced by the fraction $(t/t_{c})^{1/r}$.
This is an accelerated process of energy accumulation, which may be
repeated, by various causes, like modifications in the focal region,
an excess energy, etc. By repeating it in an infinite series, we get
a higher energy in a shorter time. It is a self-replication process,
leading to equation (\ref{17}). }

\textcolor{black}{The accumulation energy is not limited in this case,
and it may be accumulated in a finite time $t<t_{c}$; this is a self-accelerating
process of energy accumulation. Of course, infinite energies are unphysical.
From equation (\ref{17}) we get 
\begin{equation}
t=t_{0}(E_{c}/E_{0})^{r}\left(1-\frac{1}{1+E/E_{c}}\right)^{r}\label{18}
\end{equation}
 (to be compared with equation (\ref{12})) and 
\begin{equation}
\frac{t_{0}}{t}=\left(e^{-bM}+e^{-bM_{c}}\right)^{r}\,\,;\label{19}
\end{equation}
we remove the infinite energies by subtracting $e^{-\beta M_{c}}$
(corresponding to time $t_{c}$). The excedence frequency is 
\begin{equation}
P^{+}=\left(e^{-bM}+e^{-bM_{c}}\right)^{r}-e^{-\beta M_{c}}\,\,\,,\label{20}
\end{equation}
 and the density of magnitude probability is 
\begin{equation}
-dP^{+}/dM=\beta e^{-\beta M}\left[1+e^{b(M-M_{c})}\right]^{r-1}\,\,;\label{21}
\end{equation}
 the superscript \textquotedbl$+$\textquotedbl{} indicates $M>M_{c}$.
According to equation (\ref{20}), the excedence frequency takes the
value $\left(2^{r}-1\right)e^{-\beta M_{c}}$ for $M=M_{c}$, which
is smaller than the Gutenberg-Richter law, but greater than $P^{-}(M_{c})$;
this may correspond to our idea that very big earthquakes are rare
($M_{c}\gg1$ in these formulae), but not impossible. They may illustrate
the concept of \textquotedbl dragon-king\textquotedbl{} earthquakes.
Also, since all the available energy is taken by the big earthquake,
it is conceivable that small precursor earthquakes may be absent.
It is worth noting that while the Gutenberg-Richter excedence law
is only slightly modified for $M<M_{c}$ (equation (\ref{14})), and
the probability law is left unchanged (equation (\ref{15})), the
self-replication mechanism generates a different functional dependence
of these distributions for $M>M_{c}$ (equations (\ref{20}) and (\ref{21})). }

\textcolor{black}{In order to integrate the two pictures given above
we use the fact that the excedence frequency is a monotonously decreasing
function of magnitude. Consequently, we can define a crossover range
of magnitudes $M_{c}^{-}<M<M_{c}$, where $M_{c}^{-}$ is the solution
of the equation $P^{-}(M_{c}^{-})=P^{+}(M_{c})$, }\textcolor{black}{\emph{i.e.
}}\textcolor{black}{
\begin{equation}
M_{c}^{-}=M_{c}-\frac{1}{b}\ln2\,\,;\label{22}
\end{equation}
 for the reference values given above, we get $M_{c}^{-}=M_{c}-0.2$.
The Gutenberg-Richter distribution acquires this value for 
\begin{equation}
M_{c}^{+}=M_{c}-\frac{1}{\beta}\ln\left(2^{r}-1\right)\simeq M_{c}+0.24\,\,.\label{23}
\end{equation}
The difference between these two values, 
\begin{equation}
\Delta M=M_{c}^{+}-M_{c}^{-}=-\frac{1}{\beta}\ln\left(1-2^{-r}\right)\simeq0.43\label{24}
\end{equation}
 can be taken as the magnitude range of the big earthquakes (in the
vicinity of a critical magnitude $M_{c}$). This limitation removes
the unphysical region of $M$ very close to, and smaller than $M_{c}$.
On the other hand, $P^{+}$ is a slowly decreasing function of $M$,
and we may reasonably limit ourselves to $M$ close to, and larger
than $M_{c};$ in this region we may use the approximation $P^{+}\simeq\left(2^{r}-1\right)e^{-\beta M_{c}}$.
It is worth noting that in this region ($M\gtrsim M_{c}$) the excedence
frequency is reduced by the factor $2^{r}-1\simeq0.58$ in comparison
with the Gutenberg-Richter law, while de magnitude probability (equation
(\ref{21})) is reduced by the factor $2^{r-1}\simeq0.79$ (for $r=2/3$);
this reflects the idea that extremely big earthquakes, though highly
improbable, are not impossible. }

\begin{figure}
\begin{centering}
\textcolor{black}{\includegraphics[clip,scale=0.5]{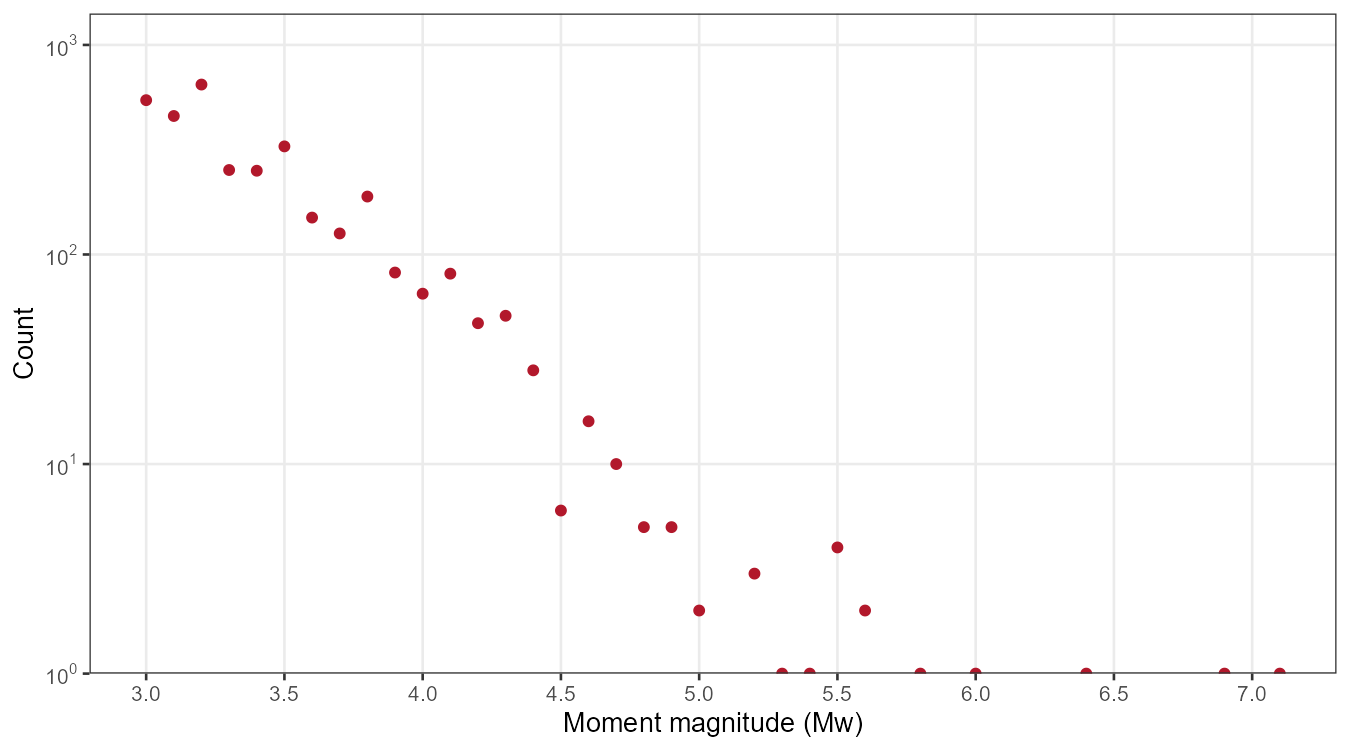}}
\par\end{centering}
\textcolor{black}{\caption{\textcolor{red}{Magnitude distribution for $3361$ Vrancea earthquakes
(1981-2018; lat. $45^{\circ}-46^{\circ}$, long. $26^{\circ}-27^{\circ}$)
with magnitude $M\protect\geq3$ ($\Delta M=0.1$). The fitting parameters
are $-\ln t_{0}=10.62$, $\beta=2.1$ (error $\simeq10\%$). Two distinct
branches on both sides of magnitude $\simeq5.5$ are present (note
the big earthquakes with $M\protect\geq6$; Roum. Earth. Catalog 2018)}.\label{fig:Fig. 3}}
}

\end{figure}

\textcolor{black}{According to the above discussion the excedence
frequency should be written as 
\begin{equation}
P=\left\{ \begin{array}{c}
e^{-\beta M}-e^{-\beta M_{c}}\,\,,\,\,0<M<M_{c}^{-}\,\,,\\
\left(2^{r}-1\right)e^{-\beta M_{c}}\,\,,\,\,M\gtrsim M_{c}\,\,.
\end{array}\right.\label{25}
\end{equation}
This function and its logarithm are shown schematically in Figs. \ref{fig:Fig.1},
\ref{fig:Fig.2}. Equation (\ref{25}) (first row) can be used to
fit the empirical data, with the fiting parameters $\beta$ (and $r$),
$t_{0}$ and $M_{c}$. It is worth noting that the magnitude distribution
of Vrancea earthquakes shown in Fig. \ref{fig:Fig. 3} seems to exhibit
two distinct branches around magnitude $M=5.5$. }

\textcolor{black}{The above calculations are valid for independent
earthquakes. If they are time-correlated, the excedence frequency
is given by 
\begin{equation}
P=\frac{2t_{0}}{t+t_{0}}\label{26}
\end{equation}
 (instead of $t_{0}/t$; we recall that $t_{0}<t<\infty$) (Apostol,
2021). By using this formula in equations (\ref{12}) and (\ref{18}),
we get the excedence frequency 
\begin{equation}
P\simeq\left\{ \begin{array}{c}
\frac{2\left(e^{-\beta M}-e^{-\beta M_{c}}\right)}{1+e^{-\beta M}}\,\,,\,\,0<M\lesssim M_{c}^{-}\,\,,\\
2\left[\left(e^{-bM}+e^{-\beta M_{c}}\right)^{r}-e^{-\beta M_{c}}\right]\simeq2\left(2^{r}-1\right)e^{-\beta M_{c}}\,\,,\,\,M\gtrsim M_{c}\,\,\,,
\end{array}\right.\label{27}
\end{equation}
 where we neglect, wherever irrelevant, the small term $e^{-\beta M_{c}}$;
the parameter $M_{c}^{-}$ remains unchanged, while $M_{c}^{+}$ decreases.
For $M\rightarrow0$ equation (\ref{27}) exhibits the roll-off effect,
where the slope of $P$ tends to $-\beta/2$ (instead of $-\beta$).
For $M\lesssim M_{c}^{-}$ and $M\gtrsim M_{c}$ the excedence frequency
and the magnitude probability of the correlated big earthquakes are
enhanced by a factor $2$ in comparison with the uncorrelated ones;
the Gutenbreg-Richter magnitude probability in this region is increased
by the factor $2\left(2^{r}-1\right)\simeq1.16$; very likely, the
correlations do not affect too much the big earthquakes. }

\textcolor{black}{In conclusion, we have described above two self-replication
mechanisms acting on the energy accumulation in a seismic focus. For
magnitudes smaller than a critical, narrow region around a magnitude
$M_{c}$, the self-replication mechanism slows down the energy acumulation,
while for magnitudes larger than that critical region, the self-replication
mechanism accelerates the energy accumulation. In both cases the excedence
frequency decreases. The Gutenberg-Richter magnitude probability is
unchanged on the left and decreases on the right of the critical region.
There exists a small cross-over in the magnitude range around $M_{c}$,
where the excedence frequency, as a well-defined function of magnitude,
does not exist, and the probability density is zero (a magnitude gap).
We cannot say why, and in what conditions, these two self-replicating
mechanism act. However, we can accept that big earthquakes are rare,
so their accumulation energy is slowed down, while excessively large
earthquakes are not forbidden, and they may suffer an accelerated
energy accumulation. This general standpoint seems to be supported
by some empirical evidence. Also, we have examined the effect of the
correlations on this anomalous behaviour of the big earthquakes, and
found out that such correlations are not likely to affect these earthquakes
very much. The two mechanisms of producing big earthquakes may bear
some relevance for the occurrence of small, precursor earthquakes,
as discussed above. }

\textbf{\textcolor{black}{Acknowledgments.}}\textcolor{black}{{} The
author is indebted to the colleagues in the Institute of Earth's Physics,
Magurele, and to the members of the Laboratory of Theoretical Physics,
Magurele, especially to dr. L. C. Cune, for many enlightening discussions.
The author acknowledges very valuable comments and suggestions from
both the anonymous Reviewers.}

\textcolor{black}{REFERENCES}

\textcolor{black}{Apostol, B. F. (2006a). A model of Seismic Focus
and Related Statistical Distributions of Earthquakes. Phys. Lett.
A357, 462-466. }

\textcolor{black}{Apostol, B.F. (2006b). Euler's transform and a generalized
Omori's law. Phys. Lett. A351, 175-176.}

\textcolor{black}{Apostol, B.F. (2021). Correlations and Bath\textquoteright s
law. Results in Geophysical Sciences 5, 100011.}

\textcolor{black}{Bhattacharya, P., Chakrabarti, C. K., Kamal, K.
D. \& Samanta. (2009). Fractal models of earthquake dynamics; in }\textcolor{black}{\emph{Reviews
of Nolinear Dynamics and Complexity}}\textcolor{black}{, H. G. Schuster,
ed., pp.107-150, NY, Wiley. }

\textcolor{black}{Frohlich, C. \& Davis, S. D. (1993). Teleseismic
$b$ values; or much ado about $1.0$, J. Geophys. Res. 98 631-644.}

\textcolor{black}{Gutenberg, B. \& Richter, C. (1944). Frequency of
earthquakes in California. Bull. Seism. Soc. Am. 34, 185-188.}

\textcolor{black}{Gutenberg, B.; Richter, C. (1956). Magnitude and
energy of earthquakes. Annali di Geofisica, 9, 1-15 (Ann. Geophys.
}\textbf{\textcolor{black}{2010}}\textcolor{black}{, 53, 7-12).}

\textcolor{black}{Hamilton, T., \& McCloskey, J. (1997). Breakdown
in power-law scaling in an analogue model of earthquake rupture and
stick-slip. Geophys. Res. Lett. 24, 465-468. }

\textcolor{black}{Hanks, T.C. \& Kanamori, H. (1979). A moment magnitude
scale. J. Geophys. Res}\textcolor{black}{\emph{. }}\textcolor{black}{84,
2348-2350.}

\textcolor{black}{Ishibe, T. \& Shimazaki, K. (2012). Characteristic
earthquake model and seismicity around late quaternary actaive faults
in Japan. Bull. Seism. Soc. Am. 102, 1041.}

\textcolor{black}{Jones, L. M. (1994). Foreshocks, aftershocks and
earthquake probabilities: accounting for the Landers earthquake. Bull.
Seism. Soc. Am. 84, 892-899.}

\textcolor{black}{Kanamori, H. (1977), The energy release in earthquakes.
J. Geophys. Res. 82, 2981-2987.}

\textcolor{black}{Lay, T. \& Wallace, T. C. (1995). }\textcolor{black}{\emph{Modern
Global Seismology}}\textcolor{black}{, Academic Press, San Diego,
California.}

\textcolor{black}{Li, J., Sornette, D., Wu, Z. \& Li, H. (2024). New
horizon in the statistical physics of earthquakes: dragon-king theory
and dragon-king earthquakes. arXiv preprint arXiv:2408.10857 {[}physics.geo-ph{]}. }

\textcolor{black}{Main, I. G. (1992). Earthquake scaling. Nature,
357, 27-28.}

\textcolor{black}{Pelletier, J. D. (2000). Spring-block models of
seismicity: review and analysis of a structurally heterogeneous model
coupled to the viscous asthenosphere; in }\textcolor{black}{\emph{Geocomplexity
and the Physics of Earthquakes}}\textcolor{black}{. vol. 120., Rundle,
J. B., Turcote, D. L., \& Klein, W., eds. NY, Am. Geophys. Union. }

\textcolor{black}{Pisarenko, V. F. \& Sornette D. (2004). Statistical
detection and characterization of a deviation from the Gutenberg-Richter
distribution above magnitude 8{[}J{]}. Pure and Applied Geophysics
161, 839-864. }

\textcolor{black}{Roum. Earth. Catalog (2018), https://www.infp.ro/index.php?i=romplus}

\textcolor{black}{Saichev, A. \& Sornette, D. (2006). Power law distribution
of seismic rates: theory and data. Eur. Phys. J. B49, 377-401.}

\textcolor{black}{Saichev, A. \& Sornette, D. (2007). Power law distribution
of seismic rates. Tectonophysics, 431, 7-13.}

\textcolor{black}{Sammis, C. G. \& Sornette, D. (2002). Positive feedback,
memory, and the predictability of earthquakes. Proc. Nat. Acad. Sci.
USA 99, (Suppl. 1) 2501-2508. }

\textcolor{black}{Sornette, D. (2009). Dragon-kings, black swans and
the prediction of crises. arXiv preprint arXiv: 0907.4290.}

\textcolor{black}{Sornette, D., Wei, X. \& Chen, X. (2026). Self-arresting
and runaway earthquakes: nucleation, propagation, Gutenberg-Richter
law and dragon-king events. arXiv preprint arXiv: 2402.14626v2 {[}physics.geo-ph{]}. }

\textcolor{black}{Stein, S., \& Wysession, M. (2003). }\textcolor{black}{\emph{An
Introduction to Seismology, Earthquakes, and Earth Structure}}\textcolor{black}{,
Blackwell, NY.}

\textcolor{black}{Udias, A. (1999). }\textcolor{black}{\emph{Principles
of Seismology}}\textcolor{black}{, Cambridge University Press, NY.}

\textcolor{black}{Utsu, T. \& Seiki, (1955). A. A relation between
the area of aftershock region and the energy of the mainshock (in
Japanese). J. Seism. Soc. Japan 7, 233-240.}

\textcolor{black}{Utsu, T. (1969). Aftershocks and earthquake statistics
(I): some parameters which characterize an aftershock sequence and
their interaction. J. Faculty of Sciences, Hokkaido Univ., Ser. VII
(Geophysics) 3, 129-195; (II) 196-266.}

\textcolor{black}{Wesnousky, S. G. (1994). The Gutenberg-Richter or
characteristic earthquake distribution, which is it? Bull. Seism.
Soc. Am. 84, 1940.}

\textbf{\textcolor{black}{Statements and Declarations}}

\textbf{\textcolor{black}{Funding}}

\textcolor{black}{This work was carried out within the Program Nucleu
SOLARISC, contract \#24N/03.01.2023, funded by the Romanian Ministry
of Research, Innovation and Digitalization, project \# PN23360202.}

\textbf{\textcolor{black}{Competing interests}}

\textcolor{black}{The author has no relevant financial or non-financial
interests to disclose. }

\textbf{\textcolor{black}{Author contributions}}

\textcolor{black}{Not applicable}
\end{document}